\documentclass[twocolumn]{aastex631}  

\usepackage{graphicx}
\usepackage{amsmath}
\usepackage{txfonts}
\usepackage{natbib}         
\usepackage{color}
\usepackage[normalem]{ulem}             
\newcommand{\kms}{km~s$^{-1}$\,}

\chardef\us=`\_

\begin{document}

   \title{The Transition Region of Solar Flare Loops}

   \author{Costis Gontikakis}
   \affiliation{Research Center for Astronomy and Applied Mathematics, Academy of Athens,\\
    Soranou Efesiou Str., 4, 11527, Athens Greece}

   \author{Spiro K. Antiochos}
   \affiliation{CLaSP, University of Michigan, Ann Arbor, MI, 48109, USA}
    \author{Peter R. Young}
  \affiliation{NASA Goddard Space Flight Center, Solar Physics Laboratory, Heliophysics Science Division, Greenbelt, MD 20771, USA}
   \date{Received ...; accepted ...}
 
  \begin{abstract}
The transition region between the Sun's corona and chromosphere is important to the mass and energy transfer from the lower atmosphere to the corona; consequently, this region has been studied intensely with ultraviolet (UV) and extreme ultraviolet observations. A major result of these studies is that the amount of plasma at temperatures $< 10^5$~K, is far too large to be compatible with the standard theory of thermal conductivity. However, it is not clear whether the disagreement lies with a problem in the observations or in the theory. We address this issue by analysing high-spatial and temporal resolution EUV observations from an X1.6-class flare taken with the {\it Interface Region Imaging Spectrograph} (IRIS) and the {\it Solar Dynamic Observatory/Atmospheric Imaging Assembly} (SDO/AIA). These data allow us to isolate the emission of flare loops from that of surrounding structures. We compare the Emission Measures (EMs) derived from the \ion{C}{2} 1334.525\AA, \ion{Si}{4} 1402.770\AA\ transition region spectral lines, the \ion{Fe}{21} 1354.066\AA\  flare line and the AIA 171\AA\ coronal images. We find that the EM ratios are incompatible with a standard conduction-dominated transition region model. Furthermore, the large increases in the EM magnitudes due to flare heating make it highly unlikely that the disagreement between data and theory is due to observational uncertainties in the source of the emission. We conclude that the standard Spitzer Harm thermal conductivity must be invalid for, at least, flare loops. We discuss the possibility that turbulent suppression of thermal conduction can account for our results.
   \end{abstract}


%

\section{Introduction}  \label{sec:intro}

A decades old challenge in solar physics has been to understand the structure of the so-called transition region (TR) that connects the chromosphere to the corona  \citep[e.g.][]{Mariska}. Beginning with the early X-ray and extreme ultraviolet (EUV) telescopes of \textit{Skylab}, the corona was observed to consist of a collection of quasi-isothermal, $T \sim 10^6$ K, loops with length scales of tens of Mm, while the chromosphere was also observed to be a relatively isothermal, $T \sim 10^4$ K, layer with thickness of a few thousand km. The TR between these two, defined as the region with temperature between roughly half that of the corona and twice that of the chromosphere, was unresolvably thin, implying that the temperature (and density) gradients there were large.

Although the temperature and density profiles of the TR could not be measured directly, their spatial variations could be tightly constrained by measuring the intensities of emission lines formed at various temperatures within the TR. For optically thin UV/XUV emission lines, it is well-known that the intensity varies as the square of the electron density, $N_e^2$, multiplied by a temperature-dependent contribution function that is usually sharply peaked at the formation temperature of the line \citep{Phillips_2008}. This implies that the intensity of a line formed at temperature $T$ is proportional to the amount of material at that temperature weighted by the density at that temperature.  Approximating the TR of a coronal loop as a 1D structure then the line intensities can, in principle, be inverted to yield the so-called Differential Emission Measure, $DEM(T)$.  Unfortunately, the $DEM(T)$ has not been consistently defined by the community. We and others \citep[e.g.][]{Dere_1979,Athay81,Antiochos_Noci_1986} prefer to define it in terms of the temperature scale height; whereas many others \citep[e.g.][]{Vesecky_1979,Klimchuk08,Del_Zanna_2015,Emslie_Bradshaw_2022} use simply the temperature gradient. Since it is trivial to convert from one form to the other, and since the latter form seems to be more commonly used in recent work, we will adopt it here, in which case the $DEM$ is given by: 
\begin{equation}
\label{eqn:DEM}
DEM = N_e^2 (dT/ds)^{-1}. 
\end{equation}

Given that the $DEM$ is the primary quantity that is derivable from the observations, a vast amount of work has conducted over the past few decades on methods for performing the inversion to derive the $DEM$ from data \citep{Dere_1979,Del_Zanna_2018, Hannah_2012}, on analysis of the data to determine the $DEM$ for observed coronal regions \citep{Brooks_2011,Mulay_2017,Graham_2013}, and on derivation of the $DEM$ from theoretical/numerical models of coronal loops \citep{Antiochos_2003, Underwood_1978}. The generic result found from this large body of work is that coronal loop theory is in good agreement with the observed $DEM$ for the so-called upper TR (temperatures between $\sim 10^5$ K and $\sim 10^6$ K for a 2MK corona), but fails badly for the lower TR, temperatures from  $\sim 10^5$ K down to chromospheric temperatures of $\sim 10^4$ K.

The reason for the failure can be seen directly from the form of the thermal conductivity for a solar coronal plasma. The simplest model for the TR thermal structure is that of a constant heat flux, $F_c$, from the coronal portion of a loop down to the chromospheric footpoint. Assuming that the heat flux is well described by the standard Spitzer-Harm collisional conductivity \citep{Spitzer_Harm_1953} yields:
\begin{equation}
\label{eqn:H_T}
    (dT/ds)^{-1} =  10^{-6}\, T^{5/2} / F_c.
\end{equation}
Equation~\ref{eqn:H_T}, although very simple, has enormous implications for the structure of the Sun's atmosphere. It immediately explains the very thin width for the TR. Assume, for example, that a coronal loop has a maximum temperature of order 2~MK, and a length of 10~Mm, then from Equation~\ref{eqn:H_T} the scale height, \textbf{$T(dT/ds)^{-1}$,} at the temperature of $10^5$~K is three and a half orders of magnitude smaller, $\sim 30$~km, which is well below the resolution of present-day instrumentation. For temperatures in the lower transition region, the scale height is even smaller. Since the TR is so thin, the pressure across it can be assumed to be close to constant; therefore, substituting Equation~\ref{eqn:H_T} into Equation~\ref{eqn:DEM} yields the key result:
\begin{equation}
\label{eqn:scaling}
    DEM(T) \sim T^{1/2} / F_c.
\end{equation}
Equation~\ref{eqn:scaling} implies that the $\rm{DEM}$ should \textit{increase}  with $T$ from the chromosphere to the corona. Observations of the upper TR are, indeed, in good agreement with this conclusion \citep{Athay_1966}, but observations of the lower TR are definitely not \citep[e.g.][]{Feldman09}. The data invariably show that the $\rm{DEM}$ decreases sharply from chromospheric temperatures to reach a minimum at $\sim 10^5$~K then it increases thereafter \citep{Gabriel_1976,Raymond_1981,Fletcher_1999}. A major caveat to Equation~\ref{eqn:scaling}, however, is that the assumption of a constant heat flux cannot be valid for the TR. As pointed out by, e.g., \citet{Vesecky_1979}, the heat flux must vanish or become small at the top of the chromosphere, otherwise the chromosphere would not be able to radiate away that energy. The boundary condition of the flux vanishing at the base of a coronal loop is central to the celebrated loop ``scaling laws"  \citep[e.g.][]{Craig_1978, Rosner_1978,  Martens_2010}.

Theoretical and numerical models have been studied by many authors including effects such as the heat flux decreasing due to the effect of radiative losses from the TR \citep[e.g.][]{Vesecky_1979} and/or enthalpy flows \citep[e.g.][]{Antiochos81}, but these invariably find that $F_c$ varies as a relatively weak power of $T$, at most linearly. Perhaps, the most comprehensive study of loop scaling laws has been carried out by \citet{Martens_2010}, who performed a detailed analysis of the dependence of loop temperature and density profiles on a broad range of factors, in particular, the form of the heating. Although the \citet{Martens_2010} study did not include determination of the $\sc{DEM}$, recent work by \citet{Emslie_Bradshaw_2022} which uses results from Martens, has shown that in general the $\sc{DEM}$   
 becomes approximately constant at lower TR temperatures. The \citet{Emslie_Bradshaw_2022} paper is especially informative, because the authors include a fairly exhaustive review of loop theory and find that a $\sc{DEM}$ that is constant or increases monotonically with $\sc{T}$ is a robust prediction of these models. These authors conclude that the disagreement with data for the lower TR remains as one of the outstanding problems in coronal physics \citep{Emslie_Bradshaw_2022}.

Generally speaking, two types of ideas have been proposed to account for the lower TR problem. They can be broadly classified as ``missing structures" or ``missing physics". The basic idea of the first type is that the models are correct, but the observations are flawed, because the spatial resolution is not sufficient to resolve the coronal loop transition region from nearby structures. Examples of such proposed structures are spicules or mini-spicules \citep{de_Pontieu_etal_2011} or small cool loops \citep{Antiochos_Noci_1986,Hansteen14}. The cool loop hypothesis is especially attractive, because theory shows that such loops cannot contribute to the $DEM$ at temperatures above $\sim 10^5$ K, exactly where the $DEM$ begins to change form. The basic idea of the second type is that the observations are correct, but standard loop models are flawed, because they are missing essential physics that invalidate the result of Equation~\ref{eqn:scaling}. Examples of such proposed effects are non-equilibrium ionization \citep{Bradshaw_Mason_2003,Spadaro_etal_1994}, strongly time-dependent heating \citep{Schmit16}, and non-collisional effects on the thermal conductivity \citep{Shoub_1987}.  

From studies of coronal loops it has proven difficult to determine which are actually at fault, the observations or the models, because the loop footpoints cannot be clearly isolated from their surroundings. The cleanest way to isolate loop TRs is by observing the so-called moss in regions of hot $T >\ 3 \times\ 10^6$ K active region loops \citep{Fletcher_1999, Berger_1999, Antiochos_2003},  When such regions are near disk center and are observed in lower temperature lines, $T<10^6$ K, the moss appears as a planar patch of bright emission. This is exactly what we would expect for the TR of such hot AR loops.  The problem is that when observed with the ultra-high resolution of IRIS, the moss appears to be highly dynamic \citep{Testa_Polito_dePontieu_2020} and breaks up into bright spots surrounded by a dark network of spicules or mini-spicules unrelated to the corona \citep{de_Pontieu_etal_2009}. These dark fibrils are undoubtedly contributing to the observations, so that isolation of the lower TR emission by itself is not possible.  

In this work we analyze the ``moss" of flare loops to attack the problem of the DEM of the lower TR, and determine whether this problem is due to missing structure or missing physics. The moss in this case is simply the bright ribbons on the chromosphere, which are the best-known and most widely-studied signature of flare energy release \citep[e.g.,][and references therein]{Qiu_J_2021}. The ribbons are the footpoints of flare loops that were formed by reconnection in a vertical current sheet as in the classic CSHKP model \citep{Carmichael_1964, Sturrock_1966, Hirayama_1974, Kopp_Pneuman_1976}. The key point of both the flare loops and the ribbons is that they exhibit a large increase in intensity, often several orders of magnitude, over pre-flare values. Therefore, by focusing on the bright ribbons we can be assured that we are measuring the TR emission of a hot flare loop rather than unrelated nearby small structures. Furthermore, the coronal temperatures of flare loops is large, typically $> 10^7$ K, so even emission from what would normally be considered coronal plasma, is actually from the TR and can be used for our analysis.  

An important issue that must first be addressed is whether the TR of a flare loop is physically similar to that of typical coronal loops in that thermal conduction plays the major role in determining the form of the DEM. Since flare heating is inherently impulsive and often dominated by heating due to electron beams, the flare TR may be very different than that of coronal loops. The evolution of flare loops has been studied in great detail by many authors \citep[e.g.,][]{Sturrock80,Fisher87,Antiochos00,Kerr20,Allred22}, however, and these analyses show that, in fact, the flare TR is conduction dominated during most of its evolution. Perhaps the most relevant study to our work here was by \citet{Antiochos00}, who measured flare ribbon formation pixel-by-pixel and with high time cadence using TRACE. These authors argued that the flare loop evolution consisted of roughly three phases, a heating phase which is expected to be fast if due to electron beams, a conduction cooling phase, and a radiative cooling phase. The observations appeared to agree well with these arguments \citep{Antiochos00}. The key point is that during the conduction dominated phase of so-called evaporative cooling \citep{Antiochos_1978}, Equation~\ref{eqn:scaling} holds, but with $F_c$ directly proportional to $T$, so that DEM(T) $\propto T^{-1/2}$, which is far too weak a decrease with $T$ to agree with observations \citep{Emslie_Bradshaw_2022}. Furthermore, during the beginning of the radiative cooling phase, which has the longest duration, conduction plays the major role in transferring heat to the loop TR, so Equation~\ref{eqn:scaling} still holds and the DEM varies as in a classic coronal loop TR. We conclude that except for a possible initial beam heating phase, a flare loop TR does provide a valid test of conduction-dominated TR theory.

\section{Observation and data analysis}  \label{sec:observ}
We analyse an X1.6-class solar flare,  observed with the {\it Interface Region Imaging Spectrograph} \citep[IRIS:][]{2014SoPh..289.2733D}  spectra and the {\it Solar Dynamic Observatory}/{\it Atmospheric Imaging Assembly} \citep[SDO/AIA:][]{Lemen_etal_2012} images. The flare took place on 2014 September 10. In X-rays, recorded with the {\it Geostationary Operational Environmental Satellite} (GOES) 1--8~\AA\ band, the flare started at 17:21~UT, had a peak intensity at 17:45~UT and showed extreme X-ray emission for more than an hour (solarmonitor.org, \citet{Gallagher_2002}) in Active Region (AR) registered as {\it National Oceanic and Atmospheric Administration} (NOAA) 12158. The solar flare was located at $(-100\arcsec, 150\arcsec )$ from the solar disk center. IRIS observed AR 12158 in a sit-and-stare mode from 11:28~UT to 17:57~UT. 
The flare ribbons crossed the IRIS slit at 17:24~UT so the IRIS spectrograph recorded the flare for the last 33 minutes of the observation. 
The principal spectral lines recorded with IRIS and used in this study are \ion{C}{2} 1343.535\AA, ({\textbf 31000 K}), \ion{Fe}{12} 1349.40\AA, ($1.6\times\ 10^6$~K), \ion{Fe}{21} 1354.066\AA, ($10^7$~K), \ion{Si}{4}, 1402.770\AA\, (80000~K), \ion{O}{4} 1399.67\AA\ and 1401.157\AA\ ($1.4\times\ 10^5$~K) where wavelengths and line formation temperatures are taken from version~10 of the CHIANTI database \citep{Dere_chianti_1997,Del_Zanna_chianti_2021}. Let us note that the accuracy of the \ion{Fe}{21} 1354.066\AA\ wavelength is lower than that of the other spectral lines \citep{Young_Tian_Jaeggli_2015} but this does not affect our results.
This solar flare is described and analysed in a number of publications \citep{Graham_2015,Li_T_and_Q_Zhang_2015, Li_D_Ning_and_J_Zhang_2015, Brannon_2015, Dudik_2016, Polito_2019, Yu_Li_etal_2020}.

\begin{figure*}
\includegraphics[width=0.7\textwidth]{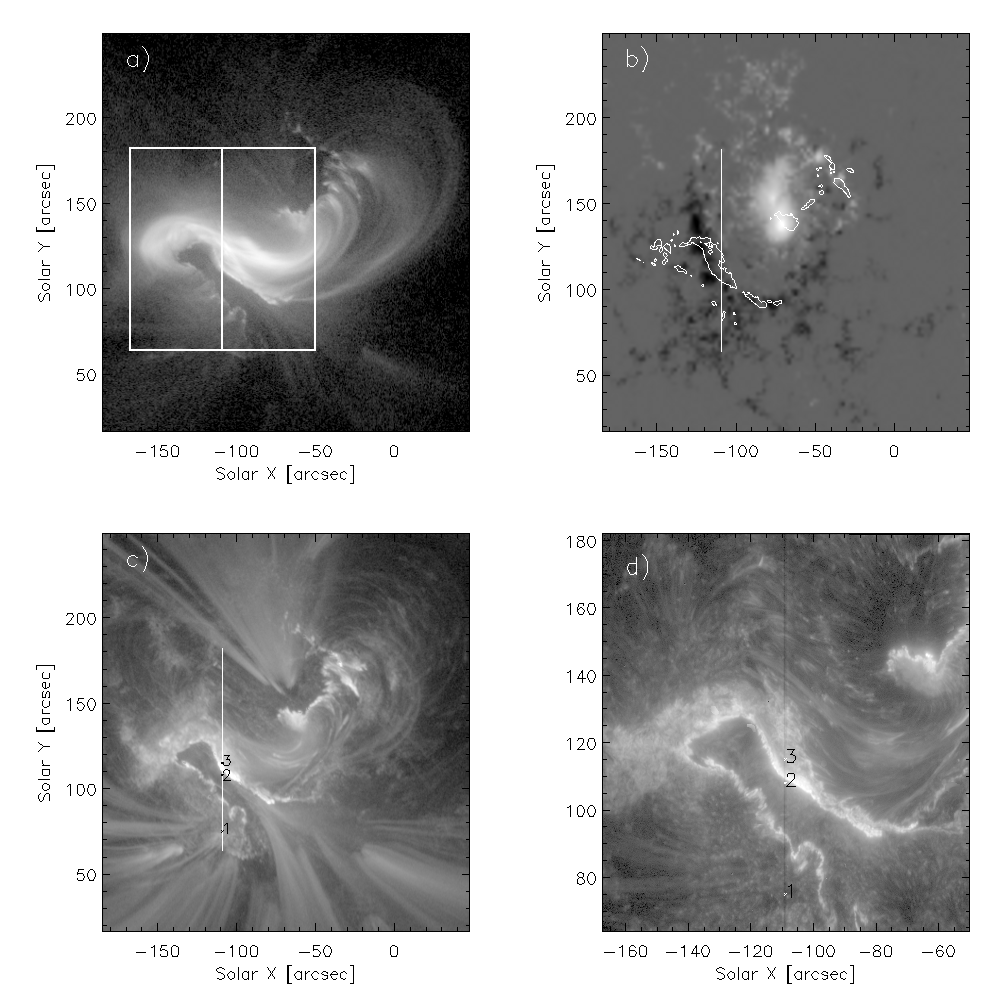}
\caption{(a): AIA 131\AA\ image of the flare loops at 17:36:36~UT, the box shows the IRIS SJI field of view (FOV), and the vertical line the  spectrograph slit. (b), HMI magnetogram observed at 16:58~UT with the same FOV as in panel a. The isophotes show the ribbons of (a). (c), the AIA 171\AA\ image of the active region at 17:36:13~UT, same FOV as in (a),(b). (d), the IRIS SJI of 1400\AA\ recorded at 17:36:40~UT. In panels c and d, the location of regions 1, 2, and 3, are indicated.}
\label{Fighmi_aia}
\end{figure*}

Figure~\ref{Fighmi_aia}a shows AR 12158 in AIA 131\AA\ dominated by hot flaring loops and ribbons at their footpoints. The field of view of the IRIS Slit Jaw Imager (SJI) is represented as a white box along with the slit position of the IRIS spectrograph. Figure~\ref{Fighmi_aia}b shows a SDO/{\it Helioseismic and Magnetic Imager} (HMI) magnetogram along with the iso-contours of the 131\AA\ flare ribbons and the IRIS slit position. The footpoints of the \lq north-west\rq\  131\AA\ hot loops  are anchored at the edges of a large sunspot with positive magnetic polarity at $\simeq$(-50\arcsec , 150\arcsec ) while the \lq south-east\rq\ footpoints are connected to complex folded ribbons centered at (-110\arcsec ,100\arcsec ) where HMI shows a negative photospheric magnetic field. The IRIS's slit crossed the South-East ribbons, as seen in Fig.~\ref{Fighmi_aia}a, at two positions (Upper and Lower positions) and recorded their temporal evolution. Figure~\ref{Fighmi_aia}c shows the AIA 171\AA\ image with the same FOV as the ones in 
Figure~\ref{Fighmi_aia}a,b. Here, loops emitted at $\simeq 10^6$~K are visible along with the ribbons. Figure~\ref{Fighmi_aia}d shows the flare ribbons as recorded in an SJI 1400\AA\ image. This image shows the emission from the \ion{Si}{4} 1393.757\AA, 1402.770\AA, spectral lines along with the continuum. The north and external part of the ribbons in Fig.~\ref{Fighmi_aia}c,d are surrounded by moss features which correspond to the footpoints of the hot loops visible in Figure~\ref{Fighmi_aia}a. In Figure~\ref{Fighmi_aia}c,d three locations, labeled 1,2 and 3 are indicated along the slit. Location 1 corresponds to a faint structure in the SJI 1400\AA\ image, and, in Fig.~\ref{Fighmi_aia}c, to a footpoint of a 171\AA\ loop. Location 2 corresponds to a ribbon area while location 3 is the footpoint of a post-flare loop. 

The evolution of post flare loops are visible in the sit-and-stare images as horizontal bright lines. We had to select a loop that is sampled by the IRIS slit at the footpoint in order to study the transition region of the loop. In many cases the loops are sampled by IRIS 
far away from their footpoints, and closer to their apex. Such loops, crossing the slit, are visible in the AIA 171\AA\ and in the 1400\AA\ slit jaw for positions $Y \gtrsim\ 140\arcsec $ from disk center, as in Figure~\ref{Fighmi_aia}d with a loop crossing the slit at $Y=143\arcsec $. In the AIA 131\AA\ images, only one or two of these loops, with their position along the slit within  $110\arcsec\ \lesssim\ Y \lesssim\ 120\arcsec $ are sampled by the IRIS slit close to their footpoint and one of them was selected as location 3. These three structures will be analysed in the following sections. 

\subsection{IRIS spectra}
In an effort to study the transition region EM we analysed mostly optically thin spectral lines recorded with IRIS. These are the \ion{C}{2} 1334.535\AA,  \ion{Si}{4} 1402.770\AA, \ion{Fe}{12} 1349.40\AA, and \ion{Fe}{21} 1354.066\AA. 
We performed Gaussian fits on the individual profiles of these spectral lines where this was possible. 
The high photon flux of  \ion{Si}{4} 1402.770\AA\ at most parts of the ribbons saturated the detector \citep[see][]{Li_T_and_Q_Zhang_2015}. Therefore, Gaussian fits were performed only on the unsaturated \ion{Si}{4} 1402\AA\ profiles. The vast majority of the unsaturated \ion{Si}{4} profiles is best described using a double Gaussian fit \citep{Brannon_2015}. 
During flares where plasma density is high, 
\ion{Si}{4} 1402.770\AA\ could present opacity effects \citep{Jeffrey_2018}. 
Important opacity effects for the \ion{Si}{4} 1402.770\AA\ are expected at the ribbons of the flares \citep{Kerr_2019}.
\ion{Si}{4} 1393.757\AA\ line is absent from our observation and we cannot check the \ion{Si}{4} 1402.770\AA\ opacity by calculating the 1393/1402 intensity ratio \citep{Gontikakis_2018}. 
Care was taken to use individual \ion{Si}{4} 1402.770\AA\  profiles that are well described by a double Gaussian fit. As reversed \ion{Si}{4} 1402.770\AA\ profiles are absent and single and double Gaussian fits are rather successful, optical thickness has a small influence for \ion{Si}{4} 1402.770\AA\  \citep{Yu_Li_etal_2020}.
\ion{C}{2} 1334.535\AA\ is an optically thick line, but is retained in our analysis as it is formed in the high chromosphere-low transition region, close to the \ion{Si}{4} 1402.770\AA\ conditions.

\ion{Fe}{21} 1354.066\AA\ is blended with the \ion{Fe}{2} 1352.69\AA, 1354.06\AA, and \ion{C}{1} 1354.28\AA\ lines. During chromospheric evaporation the \ion{Fe}{21} 1354.066\AA, line shows blueshifts of $\simeq\ -250$~\kms\ \citep{Graham_2015} that correspond to a shift of $\simeq$1.3\AA. In this wavelength range,  the \ion{Fe}{21} is blended with many \ion{Fe}{2}, \ion{Si}{2} and H$_2$ lines that are taken into account (see \citet{Li_D_Ning_and_J_Zhang_2015}). 

\begin{figure*}
\includegraphics[width=0.7\textwidth]{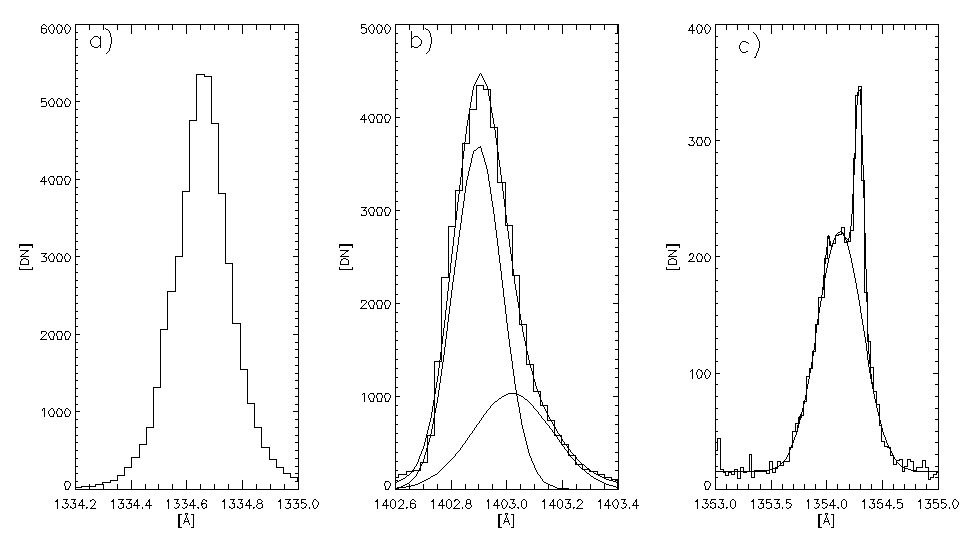}
\caption{Individual profiles of \ion{C}{2} 1334.535\AA, (a) \ion{Si}{4} 1402.770\AA, (b) and \ion{Fe}{21} 1354.066\AA\ (c) along a coronal loop. Histogram lines represent the data and, in panels b,c, continuous lines represent the Gaussian fits. The profiles are recorded along a post-flare loop on solar Y=123\arcsec\  along the slit, at 16:40:36 UT.} 
\label{Figure3_profiles}
\end{figure*}

\ion{Fe}{12} 1349.40\AA\ is a very faint forbidden line in the IRIS spectrum \citep{Testa_2016}. Although Gaussian fits were possible at some locations, in all cases  the line width was lower than the thermal width of the \ion{Fe}{12} line \citep{Guglielmino_2019}, suggesting the line belonged to a cooler species. Therefore this line was not used in our analysis.
Figure~\ref{Figure3_profiles} shows individual profiles of the \ion{C}{2} 1334.535\AA, \ion{Si}{4} 1402.770\AA, and \ion{Fe}{21} 1354.066\AA\ recorded along a post-flare loop at solar Y=123\arcsec, along the slit, at 16:40:36~UT. In Fig.~\ref{Figure3_profiles}a the \ion{Si}{4} 1402.770\AA\ is described with a double Gaussian fit. In Fig.~\ref{Figure3_profiles}b. The \ion{Fe}{21} 1354.066\AA\ is bright and very broad while a second Gaussian component is used to describe the \ion{C}{1} 1354.28\AA, line. We used a single Gaussian function to describe the \ion{Fe}{21} 1354.066~\AA\ line.
\ion{O}{4} 1399.766\AA, 1401.157\AA, are forbidden lines that can be used to measure the electron density \citep{Young_2015}. Details on the study of the \ion{O}{4} lines will be given in a later section.

\subsection{AIA filtergrams}
The AIA images  are affected by saturation caused by the bright emission of the flare. However, during the flare, AIA records a series of images with lower exposure times than the nominal $\simeq$ 2~s exposure time \citep{Lemen_etal_2012}.
We selected these low exposure images where the saturation effects are reduced. We defined the projection of the IRIS slit on sequences of AIA images in 171\AA\ and 131\AA. Then we selected the slit sections from the AIA images to create \lq sit and stare like\lq\ AIA images similar to the IRIS spectral images \citep{Li_T_and_Q_Zhang_2015}.
For this we carefully aligned the AIA images with the IRIS spectral images. Moreover, we binned the IRIS intensities along the slit to the AIA spatial resolution. Finally, we interpolated in time the sliced images that were affected by saturation. 
\begin{figure*}
\includegraphics[width=0.7\textwidth]{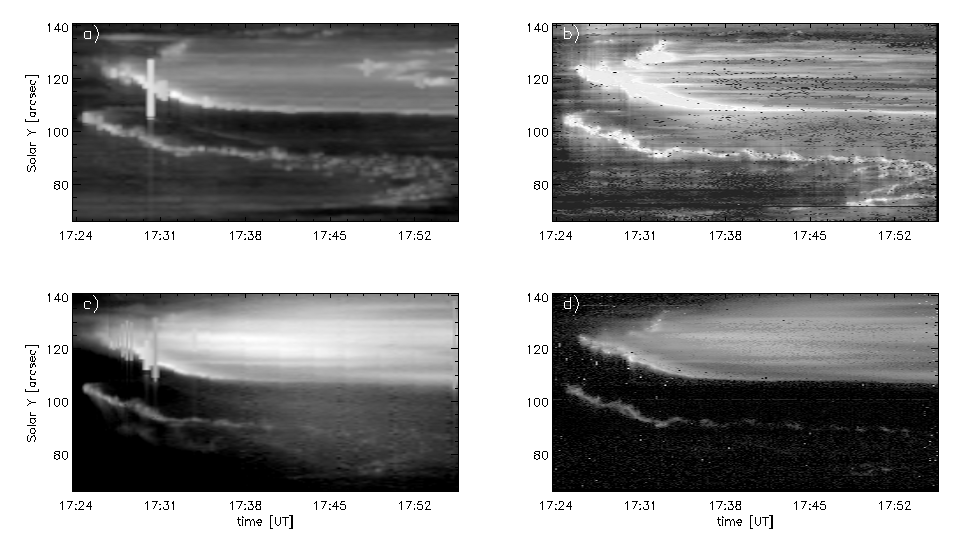}
\caption{(a): AIA 171\AA\ sit and stare like image showing times (horizontal axis) from 17:23:42~UT to 17:57:59~UT. The vertical axis
shows structures co-spatial to a section of the IRIS slit.
(b): Intensity map from the Gaussian fit on the \ion{Si}{4} 1402\AA\ first spectral component. Intensities along the upper ribbon are saturated up to 21.7~min. (c): AIA 131\AA\ sit and stare like images as in (a). (d): Intensity map derived from Gaussian fit on the \ion{Fe}{21} 1354.066\AA\ line. Note that the structures along the low ribbon may include cool emission as the Gaussian fit of the \ion{Fe}{21} line is very uncertain.}
\label{Figure_sit_and_stare}
\end{figure*}

Figure~\ref{Figure_sit_and_stare} shows the sit and stare images created using the AIA 171\AA\ and 131\AA\ images (panels a and c, respectively)  and the IRIS \ion{Si}{4} 1402.770\AA\ and \ion{Fe}{21} 1354.066\AA\ lines (panels b and d, respectively). The latter images show the integrated intensities derived from the Gaussian fits to the line profiles.
The images show the traces of the two upper and lower sections of the  South-East ribbon. 
The Upper section is followed by horizontal features that are
absent from the lower ribbon section. 
Comparing Fig.~\ref{Fighmi_aia} with Fig.~\ref{Figure_sit_and_stare} 
it is deduced that the horizontal features from the hot lines (Fig.~\ref{Figure_sit_and_stare}c,d) correspond to sections across the hot loops while the horizontal features in the transition region images (Fig.~\ref{Figure_sit_and_stare}a,b) correspond to the temporal evolution of the moss (see Fig.~\ref{Fighmi_aia}c,d).

\section{Calculation of electron density}
Knowledge of the electron density is important in order to have a valid $G(T,n_e)$ value for the forbidden lines. 
Measurements of electron densities from the present flare observation, using the ratio of the \ion{O}{4} 1399.766\AA\ over \ion{O}{4} 1401.157\AA\ lines, can be found in \citet{Young_2015}.  There, the \ion{O}{4} profiles, are taken on the ribbons and on a loop structure that precedes the flare. The \ion{O}{4} 1399.766\AA, 1401.157\AA\ appear bright, without blends. The 
upper ribbons, at 17:31:41~UT and at 17:41:04~UT have electron  densities of $1.04\times\ 10^{12}$~cm$^{-3}$ and $7.2\times\ 10^{11}$~cm$^{-3}$ respectively. These measurements are sampled 
in parts of the ribbons where most, if not all, the \ion{Si}{4} 1402.770\AA\ profiles are saturated.
Later during the flare, when the \ion{Si}{4} 1402.770\AA\ line profiles are not saturated at the ribbons nor at the loop locations, the \ion{O}{4} 1399.766\AA\ profile is blended with a line at 1399.7\AA, possibly caused by $H_2$ \citep{Young_2015}, along with the well separated \ion{Fe}{2} 1399.96\AA. The measured 1399/1401 ratios are in most cases higher than the theoretical prediction while the error bars are too high so that one cannot obtain electron density estimate.

For this reason we follow the method of \citet{Tian_2016} for estimating the electron density by first computing the DEM from AIA images.
Using the six AIA EUV filtergrams recorded during the flare (94\AA, 131\AA, 171\AA, 193\AA, 211\AA, 335\AA ), we calculated the DEM using the \cite{Hannah_2012} method. 
\begin{figure*}
\includegraphics[width=0.9\textwidth]{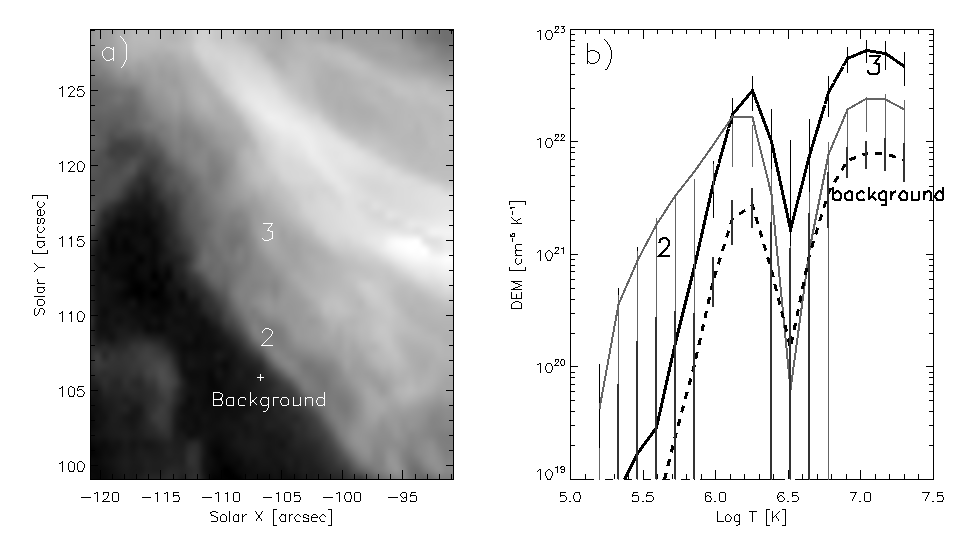}
\caption{Left: A map of the sum of the DEM of a small FOV of the active region calculated during 17:49~UT. The loop 3 and ribbon 2 are indicated on the image at the position they are observed with IRIS. A point close to the ribbon is selected as a background.
Right: The DEM calculated for the loop and ribbon and background at that time. }
\label{Figure_dem}
\end{figure*}

The DEM was calculated for a small FOV of the AIA images around the section of the IRIS slit with the selected loop  (region 3) and a section of the ribbon (region 2). The DEM calculation was performed 20 different times, where the 6 EUV AIA images are not saturated. Moreover we avoided the images for which the corresponding IRIS \ion{Si}{4} 1402.770\AA\ line is saturated on the ribbon. Figure~\ref{Figure_dem}a shows the map of the sum of the DEM 
calculated from AIA images recorded on 17:49~UT. The positions of loop 3 and ribbon 2 as well as a background point are indicated on the image.  Figure~\ref{Figure_dem}b shows the DEM curve of the loop (black line) of the ribbon (gray line) and a background point (dashed line). The ribbon DEM is higher than the loop DEM by a factor of 10 at  $\log_{10}\,T\simeq 5.0$. At $\log_{10}\,T\simeq 7.0$, however, the loop DEM is higher than the ribbon DEM by a factor of two. 
The low AIA emission on the ribbon for high temperatures is consistent with the absence of \ion{Fe}{21} emission on the ribbons. The background DEM presents also two thermal components at $log_{10} T\simeq\  $ 6.2 and 7. These background components are lower by factors 10 and 3 from the ribbon components, respectively. The $10^7$~K background component may be a scattering light effect. We performed the DEM calculation using also the routine xrt\_dem\_iterative2.pro presented in \citet{Weber_2004} finding similar results. The form of the DEM curves in Figure~\ref{Figure_dem} is quite similar to that found in recent work by \citep{del_Zanna_2022}.
The error bars of the DEM are calculated assuming an error of 20\% for the intensities. 
From \citet{Tian_2016} we have,
\begin{equation}
\mathrm{EM(AIA)} = \int  \mathrm{DEM}(T)\mathrm{d}T\,=\,f\,n_e^2\,L,
\label{integr_dem}
\end{equation}
where $f$ is the filling factor of the observed structure and $L$ the length of the structure along the line of sight. For both the ribbon and loop observations we assume that $L$ equals  the width of the structures as they appear on the images. Assuming $f=1$, we then have 
\begin{equation}
n_e\,=\,\sqrt{\frac{{\mathrm{EM(AIA)}}}{L}}
\label{electron_dens}
\end{equation}
The widths of the observed features are around 2.5 AIA pixels, and so $L=1.1\times 10^{8}$~cm. We then obtain $n_e\simeq 9\times\ 10^{10}$~cm$^{-3}$ for the loop and $n_e\simeq\, 5\times\ 10^{10}$~cm$^{-3}$ for the ribbon. We will address the filling factor assumption in Section~5.

\section{Calculation of EM}
We computed the EMs of the \ion{C}{2} 1334.535\AA,  \ion{Si}{4} 1402.770\AA, and \ion{Fe}{21} 1354.066\AA\ spectral lines and of the AIA 171\AA, 131\AA\ filtergrams. 
Using the Gaussian fits, we calculated the total intensities for each individual profile in (DN) units and transformed them to {\rm erg\, s$^{-1}$ sr$^{-1}$ cm$^{-2}$}  units, using the iris\_spec\_calib routine according to the IRIS spectrograph calibration, and using IRIS's {\it version~6} calibration, as described in \citet{2014SoPh..289.2733D}. 
To compute the EM we used the contribution functions, $G(T,n_e)$, from the CHIANTI database, which were calculated under the assumption of ionization equilibrium along with the coronal abundances from \citet{Feldman_1992}. The choice of a coronal abundance will be discussed later. 
For the IRIS spectral lines, we calculated the EMs according to: 
\begin{equation}
{\rm EM} \,=\, \frac{\mathrm{I}}{G(T_\mathrm{max},n_e)}, 
\label{eq_EM}
\end{equation}
in [cm$^{-5}]$ units, where $I$ is the intensity in physical units and $ T_\mathrm{max}$ is the maximum formation temperature of the ions. We also performed EM calculations using the CHIANTI integral\_calc routine where the contribution function is averaged over $\Delta log_{10}T = \pm 0.15$, from $T_{max}$, \citep{2010A&A...518A..42T}. The EMs calculated using the averaged $\bar G(T,n_e)$ are higher by 20\% to 30\% in comparison with the EM calculated using the maximum $G(T,n_e)$ value.

\begin{figure*}[!ht]
\includegraphics[width=1.\textwidth]{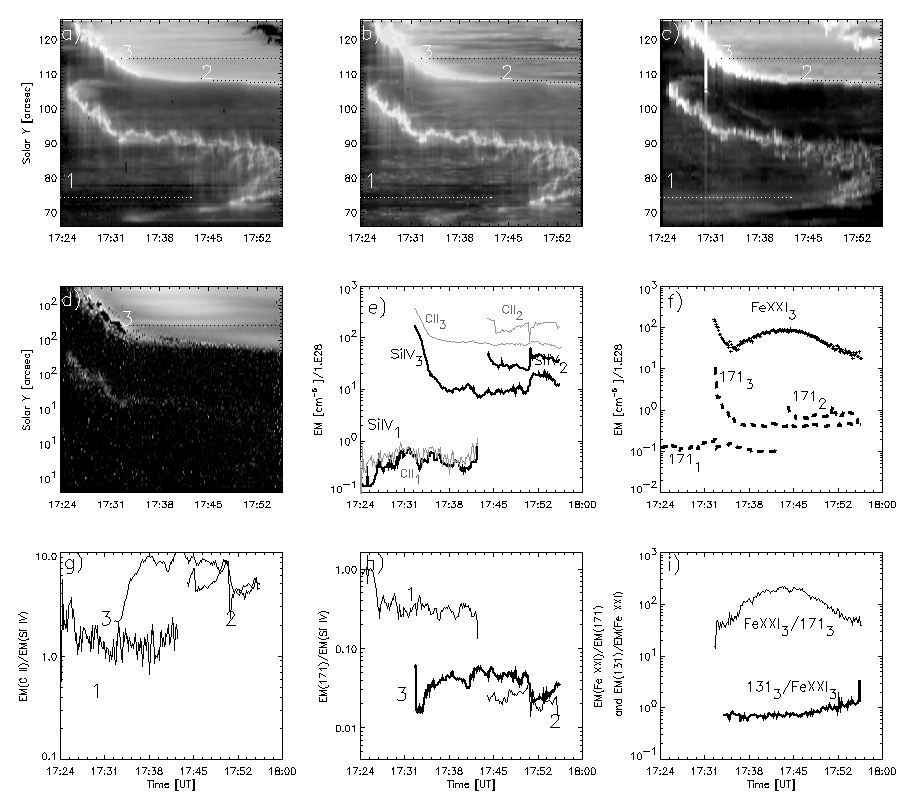}
\caption{Panels (a)-(d): intensity maps of \ion{C}{2}  1334.535\AA, \ion{Si}{4} 1402.770\AA, AIA 171\AA, and \ion{Fe}{21} 1354.066\AA\ respectively. The horizontal dotted lines show the sampling position of a region outside the flare (region 1, y=75\arcsec, time less than 17:45~UT), a ribbon (region 2, at y = 108\arcsec\  at times between 17:45~UT and 18:00~UT) and a loop (region 3, y=115\arcsec\ at times between 10:31~UT and 18:00~UT).
(e): EM of \ion{C}{2} \ 1334.535\AA\ and \ion{Si}{4} 1402.770\AA, along the three structures. (f): EM of AIA 171\AA, along the three regions and EMs of \ion{Fe}{21} 1354.066\AA\  along region 3. (g): EM ratio of \ion{C}{2} 1334.535\AA, over \ion{Si}{4} 1402.770\AA. (h), 
EM ratio of the AIA 171\AA\ over \ion{Si}{4} 1402.770\AA. (i): EM ratio of \ion{Fe}{21} 1354.066\AA\ over AIA 171\AA\ and EM ratio of AIA 131\AA\ over \ion{Fe}{21} 1354.066\AA. Both ratios are measured along region 3.}
\label{Figure_em_ratios}
\end{figure*}

\ion{Si}{4} 1402.770\AA\ is a resonance line for which the contribution function does not depend on the electron density so one can express it as $G_{1402}(T)$. However,  \ion{Fe}{21} 1354.066\AA\ is a forbidden line and $G(T,n_e)$ decreases with density above around $10^{10}$~cm$^{-3}$, hence it is important to calculate the density. For the ribbon region 3 we use $n_e\,=\, 5\times\ 10^{10}$~cm$^{-3}$, derived in the previous section.

The AIA 171\AA\ intensities are expressed in ${\rm DN\, px^{-1}\, s^{-1}}$. These are changed to emission measures in ${\rm cm^{-5}}$  units when divided with the response function of the 171\AA\ filtergram. For the calculation of the Emission Measure we followed the procedure of \citet{Li_Innes_2016}.

\ion{C}{2} 1334.535\AA\ is optically thick for most solar structures \citep{Rathore_etal_2015} but has previously been included in Emission Measure and DEM analyses \citep{Raymond_1981, Warren_etal_2001}. In \citet{Parenti_2019}, the authors restrained from the use of \ion{C}{2} 1334.535\AA\ in a DEM calculation even if its intensity is compatible with the calculation's result.
In our study, the analyzed \ion{C}{2} 1334.535\AA\ profiles are single peaked indicating that the source function of the line is rising with height in the atmosphere and that the highest layers of emitting plasma are contributing most of the photons \citep{Rathore_Carlsson_2015}. 

Figure~\ref{Fighmi_aia}c and (d) indicate the three selected regions. Region 1 is a faint structure, as seen in \ion{C}{2} 1334.535\AA\ and \ion{Si}{4} 1402.770\AA\ lines, outside the flare, which is at the footpoint of a AIA 171\AA\ loop (see Figure~\ref{Fighmi_aia}c). The results for this region serve as a test of our observations and analysis, because they should be similar to EM calculations of previous studies as we discuss in the Section~5.

Region 2 is a section of the upper ribbon. We sampled the ribbon after 17:45:27~UT which corresponds to 21.7~min after 17:23~UT which is roughly the time when the ribbons cross the IRIS FOV. Before 17:45:27~UT, the \ion{Si}{4} line is saturated at the ribbon. In any case, it is necessary to measure the ribbon properties some time after the initial energy input phase, so that a thermal conduction driven TR can be established. We estimate that the time of our measurements of region 2 should correspond to the evaporative cooling phase of a flare loop \citep{Antiochos_1978}. Location~3 corresponds to a position along the leg of a flare loop, and not solely the footpoint. Post-flare loops are formed as extensions of the upper ribbon, and on the sit and stare IRIS images appear as bright structures recorded for several minutes at the same slit position. The emission from Location~3, therefore, is a combination of coronal and TR.
As the studied emission lines are optically thin, the background emission is added to the structures. We assumed that a background emission will be similar to the emission found between the ribbons, for positions between 95\arcsec\ and 105\arcsec\ along the slit. For the \ion{Si}{4} 1402.770\AA, the background intensity is lower than the intensity of structure 3 by a factor greater than five. Therefore  we neglected the background correction in our study.

\section{Results}
Figure~\ref{Figure_em_ratios} summarizes the key results of our study.
Figures~\ref{Figure_em_ratios}(a) to (d) show the \ion{C}{2} 1334.535\AA,  \ion{Si}{4} 1402.770\AA, AIA 171\AA\ and \ion{Fe}{21} 1354.066\AA\ intensity images respectively with the three locations indicated with straight lines. Figure~\ref{Figure_em_ratios}e shows the chromospheric and Transition Region EM at the three locations as a function of time. The EMs calculated from \ion{C}{2} 1334.535\AA, (the gray lines) and  \ion{Si}{4} 1402.770\AA\ (thick black lines). Figure~\ref{Figure_em_ratios}f shows the coronal ones, the EM of  AIA 171\AA\ (dashed lines) for the three structures and \ion{Fe}{21} 1354.066\AA\ (thick black line) for region 3. The \ion{Fe}{21} 1354.066\AA\ line cannot be measured accurately outside the hottest flare emission, and nor on the ribbon area, so we omitted its EM from regions 1, and 2.
Figures~\ref{Figure_em_ratios}(g),(h) show the EM  ratios of EM(\ion{C}{2})/EM(\ion{Si}{4}), 
and EM(171)/EM(\ion{Si}{4}) respectively while Figure~\ref{Figure_em_ratios}(i) shows the ratios EM(\ion{Fe}{21})/EM(171) and  EM(131)/EM(\ion{Fe}{21}). 

In Figure~\ref{Figure_em_ratios}(e), EM(\ion{Si}{4}) and EM(\ion{C}{2}) 
are similar in region~1 with values of the order of $10^{28}$~cm$^{-5}$. In regions 2 and 3 EM(\ion{Si}{4}) is higher when compared with region~1, reaching values from $2\times\ 10^{29}$~cm$^{-5}$ to $2\times\ 10^{30}$~cm$^{-5}$.  EM(\ion{C}{2})/EM(\ion{Si}{4}) in regions 2 and 3 takes values in the range 2 to 10. There are two key points here. First, it is clear that the TR emission in the flare ribbons and loops (regions 2 and 3) is orders of magnitude larger than the non-flare emission of region~1. Therefore, the intensities we are measuring are all due to the TR of flare loops and not to some neighboring unrelated structures. Second, we note that in region~1,  \ion{C}{2} formed at approximately 31,000~K, has the same EM as  \ion{Si}{4} formed at roughly three times larger $T$. In this non-flaring TR, the DEM curve has flattened out at these low temperatures. The situation is seen to be quite different for the flare TR. The EM of the lower temperatures is an order of magnitude or so higher than at the upper temperatures; hence, the DEM is decreasing with $T$, which is contrary to what thermal conduction driven models predict.

In Figure~\ref{Figure_em_ratios}(f), EM(171)  has values of $10^{27}$~cm$^{-5}$ in region 1 and values between $3\times\ 10^{27}$~cm$^{-5}$ to $4\times\ 10^{29}$~cm$^{-5}$ in regions 2 and 3. EM(171)/EM(\ion{Si}{4}) is around 0.01--0.07 for regions 2 and 3, but around 0.3 for region 1, i.e., about an order of magnitude higher. We examine below (section 5.1 and 5.2) several possible effects that may be responsible for the reduction of EM(171) compared to EM(\ion{Si}{4}).

In region 2, the ribbon area, EM(171) is higher relative to region 1  by a factor of 6. On the other hand, the EM(\ion{Si}{4}) is higher by a factor of  $\simeq$ 100. Regions 2 and 3 have similar values of EM(171). Note that direct comparison between the  region~1 EM(171) value and  the values from regions 2 and 3 is not possible. For the latter, the emission is clearly from the TR, because the flare loops have a much hotter coronal temperature, of order $10^7$ K, but for region~1 the emission is coming from the corona. For all three regions the DEM continues to decrease with temperature, but for region~1, it is possible to argue that this is due to the contribution from low-lying cool structures. Unless one argues that these low-lying structures somehow get energized in synchrony with the flare loops, which would be in direct conflict with the standard model of flare energy release via reconnection in a coronal current sheet, then again the emission in regions~2 and 3 is solely from the TR and the DEM disagrees with theory.
EM(\ion{Fe}{21}) in region 3 is in the range $1.1\times\ 10^{30}$ to $3.7 \times\ 10^{30}$~cm$^{-5}$, so that the DEM appears to increase for temperatures above $\sim 10^6$ K, which is in agreement with theory. The filling factor and the density of region~3 in the \ion{Fe}{21} 1354.066\AA\  can be addressed here. From the intensity variation across the region~3 at different times (17:36~UT to 17:44~UT), the FWHM of the loop was found from 7 to 13 pixels along the slit. For the electron density of $10^{11}$~cm$^{-3}$ and the measured EM, the filling factor was found close to 1. Therefore, our assumption, in section~3, of a filling factor equal to 1 seems consistent. 

In Figure~\ref{Figure_em_ratios}i, EM(131)/EM(\ion{Fe}{21}) $\simeq\ 0.75$ until 17:48~UT, while it becomes higher than 1 later. The AIA 131\AA\ measurement includes emission from \ion{Fe}{21} 128.75\AA\ along with continuum and cool \ion{Fe}{8} lines. The EM comparison of AIA 131\AA, and the IRIS \ion{Fe}{21} 1354.066\AA\ line should indicate the fraction of flare emission in the AIA 131\AA\ intensity \citep{Li_Innes_2016}. In cases where AIA 131\AA\ is dominated by $10^7$~K plasma, such as in post flare loops,  a ratio EM(131)/EM(\ion{Fe}{21}) $<1$ may be explained by the different spatial resolution of AIA and IRIS along with the filling factor. 
Assuming that the IRIS filling factor equals 1 while the AIA filling factor is smaller than 1, then the ratio EM(131)/EM(\ion{Fe}{21}) is roughly equal to the AIA filling factor. 
This argument implies that the filling factor of the AIA 131\AA\ equals to $f\simeq 0.75$. Errors in the co-alignement between AIA 131\AA\ and IRIS images can also influence the ratio EM(131)/EM(\ion{Fe}{21}) (see also \citet{Li_Innes_2016}). When the ratio EM 131\AA /EM \ion{Fe}{21} becomes higher than one, cool plasma may become non-negligible in AIA 131\AA.
As this filling factor is very close to 1 we ignored its influence in our results. We refer the reader to the work of \citet{Young_doschek_2013} for EM calculations from the EIS spectrograph and AIA images and for a similar treatment of differing spatial resolutions.
\begin{figure*}[!ht]
\includegraphics[width=0.7\textwidth]{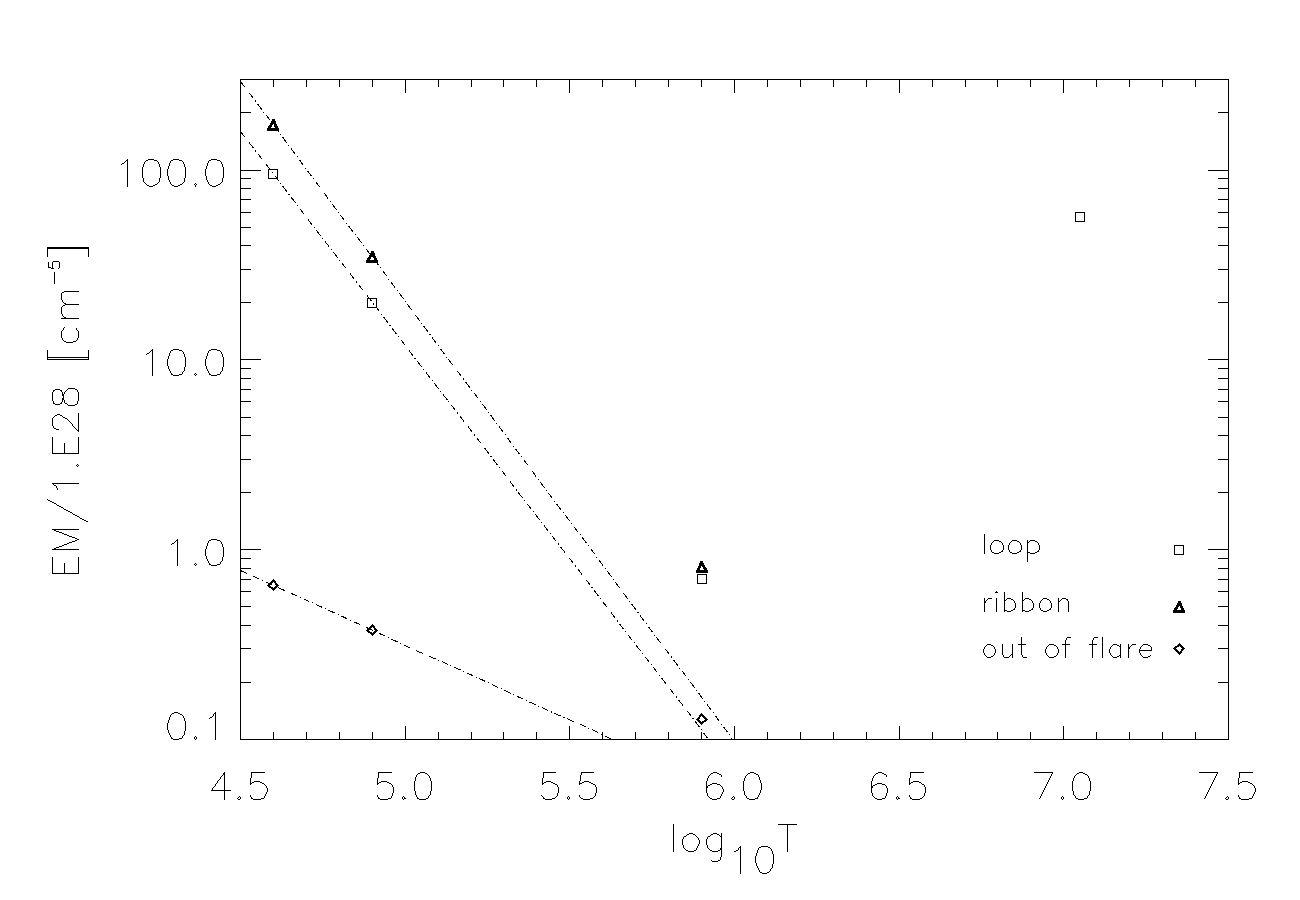}
\caption{Emission Measure values, averaged over time, as a function of temperature. Diamonds, squares and triangles represent EM from regions 1, 2 and 3 respectively. For reference, straight lines are drawn through the two values in the lower transition region.}
\label{EMasafuncoftemp}
\end{figure*}

Figure~\ref{Figure_em_ratios}(i), shows EM(\ion{Fe}{21})/EM(171)  for region 3, which is of the order of 30 to 200. This result is well within the range of DEM slopes observed for flare and hot active region loops, generally DEM $\propto T^\alpha$, with $\alpha$ ranging between 1 and 2 or so \citep{Graham_2013}. 

Figure~\ref{EMasafuncoftemp} shows the EM values shown in Figure~\ref{Figure_em_ratios}(e) and (f), averaged over time, as a function of the line formation temperature. The diamonds, squares and triangles represent regions 1 to 3 respectively, and straight-line fits to the values in the lower TR for all three regions are drawn. Clearly the DEM in the lower TR is neither increasing nor constant with temperature, in disagreement with the standard loop models \citep{Emslie_Bradshaw_2022}. We note that in all three cases the lines pass below the corresponding $DEM$ value at $10^6$ K, implying that the $DEM$ curves turn up somewhere between $10^5$ and $10^6$ K and have a minimum there. Since our study includes EM values for only four spectral lines, the plots are grossly under-sampled, especially for $T \simeq\ 10^5$~K where one expects the minimum to occur \citep{Dere_1979}. Note also that only the flare loop has emission at $10^7$~K, so that we have more than one point in the high temperature region of the $DEM$ curve. Connecting this point with the corresponding value at $10^6$ K yields a fairly strong temperature dependence, $DEM \sim T^2$. This result is in line with many other observations, which find a steeper temperature dependence of the $DEM$ in flare and hot active region loops than the standard $ \sim T$ result, but the result is consistent with loop models \citep[e.g.][]{Underwood_1978}. But as argued in \citep{Emslie_Bradshaw_2022}, the steep drop in $DEM$ with $T$ at the lower temperatures is definitely not consistent with any of the standard loop models.

On the whole, our results are in line with previous EM measurements. In \citet{Del_Zanna_2018}, Fig. 55 shows the Emission Measure loci, derived from solar irradiance measurements taken from \citet{Judge_etal_1995} calculated with updated atomic parameters. There, the EM at $\log\,T=5.9$ over the EM of \ion{Si}{4}  corresponds to 0.1 consistent with region~1 of our work. Indeed region~1 corresponds to the fainter \ion{Si}{4} 1402.770\AA\ intensity of our data set. 
For regions 2 and 3, the ratios 171\AA\ / \ion{Si}{4} 1402.770\AA\ are lower than in region~1 by a factor of $\simeq$ 10. In \citet{Young_doschek_2013}, the emission measure calculated over a flare kernel shows higher values for $\log\,T$=5.5 relative to $\log\,T$=6, by 50\%.
In \citet{Vourlidas_etal_2001}, the emission measure of the hydrogen Ly$\alpha$  is compared with the emission measure of  171\AA\ for a moss area. There, the emission measure ratio of Ly$\alpha$ over the 171\AA\ is from 0.0025 to 0.01. This result is consistent  with our calculated emission measure ratio of 171\AA\ over the \ion{C}{2} line, which is of the order of 0.005 as the formation temperature of \ion{C}{2}, is $2.5 \times\ 10^4$~K, according to CHIANTI. 

The primary result of our analysis of these flare data, is that the  emission measure ratios of AIA 171\AA\ over \ion{Si}{4} 1402.770\AA\  indicate that the observed plasma temperature and density structure for temperatures between 80\,000~K and  $\simeq\, 10^6$~K are in direct conflict with models based on classical thermal conduction. On the other hand, the EM ratio of \ion{Fe}{21} 1354.066\AA\ over AIA 171\AA\ indicates that classical thermal conduction may well be determining the high temperature transition region between $\simeq\, 10^6$~K and $5\times\ 10^7$~K.
In the following subsections, we discuss possible caveats to our results on the emission measure ratio of \ion{Si}{4} 1402.770\AA\ over AIA 171\AA\ value.

\subsection{Dielectronic Recombination Suppression}
\ion{Si}{4} belongs to the Na-like isoelectronic sequence. Theoretical intensities from spectral lines emitted by Na-like and  Li-like ions  show discrepancies when compared with observations \citep{Del_Zanna_2002}. The \ion{Si}{4} 1402.770\AA\ spectral line intensities are underestimated by factors of 2 to 3 when compared with the intensities of spectral lines formed at the same temperature, and emitted by ions of different isoelectric sequences.
These deviations could be caused by the dependence of dielectronic recombination on electron collisions which are not taken into account in the coronal approximation. We used the \citet{Nikoli__2013} description of the Dielectronic Recombination Suppression (DRS) implemented in the CHIANTI software \citep{Del_Zanna_chianti_2021}. 

We computed the contribution function G$_{1402}$(T) of the \ion{Si}{4}  1402.770\AA\ line taking into account the DRS. The \ion{Si}{4}  formation temperature decreases from 79000K to 70000K when we compare the
G$_{si4}$(T), computed under coronal approximation, with the G$_{1402}$(T) with the DRS effect. Moreover, for electron densities of $10^{10}$ and $10^{12}$~cm$^{-3}$, the G$_{1402}$(T) rises by 1.58 and 1.75 relative to the coronal approximation contribution function.
Therefore, the DRS dependence on electron density, according to \citet{Nikoli__2013}, cannot explain the variation of the Emission Measure ratios of a factor 10.

\subsection{Element Abundances}
Element abundances in the corona are known to differ from those in the photosphere, with the differences correlated with the element's first ionization potential (FIP)---see, e.g., \citet{Feldman_1992b} and \citet{2015LRSP...12....2L}. The ratio of the coronal abundance to the photospheric abundance is referred to as the FIP bias, and it is found to vary in different coronal features. Generally low-FIP elements (FIP $< 12$~eV) are enhanced in the corona relative to high-FIP elements.
The uncertainty remains in the value of the FIP bias for flare ribbons \citep[e.g.,][]{Kerr_2019,Graham_2013, Young_doschek_2013}. 

The AIA 131 and 171~\AA\ channels are dominated by iron lines \citep{O_Dwyer_etal_2010}, hence we expect their emission to behave as a low-FIP species. There is a continuum contribution during flares that will be dominated by hydrogen, a high-FIP element. However, we find that the continuum contributes at most 15\%\ to the two AIA channels and so we ignore this effect here.
Silicon is also a low-FIP element and so we do not expect ratios of \ion{Si}{4} to the AIA channels to be affected by the FIP effect. 

Carbon is classed as a high-FIP element, however, and so the EM(\ion{C}{2})/EM(\ion{Si}{4}) ratio (Figure~\ref{Figure_em_ratios}g) is sensitive to the choice of element abundances. If photospheric abundances had been used, the ratios would decrease by a factor three.

\subsection{Non-Maxwellian Distribution Functions}
The final effect that may be influencing the interpretation of our observation is that the particle distributions may be non-Maxwellian, which could change substantially the UV and EUV emission lines properties, such as the formation temperature, contribution function, etc. 
Many measurements indicate that particle distributions in the solar transition region and corona diverge from a Maxwellian function at high kinetic energies, and can be described more accurately as $\kappa$-distributions \citep{Dudik_2017}. This would be especially true in the onset phase of a flare where, before significant evaporation occurs, there is an explosive energy release in a relatively low-density plasma. For the transition region, a highly non-Maxwellian plasma, corresponding to a low $\kappa$ value, will reduce the formation temperature of the \ion{Si}{4} ion to lower values (logT$\simeq$ 4.8 - 4.6), which will enhance the emission measure \citep{Dudik_2014}. Our results, therefore, could be influenced by non-Maxwellian distributions. It should be noted, however, that we study the emission 30 minutes after the start of the flare, and at these times non-equilibrium ionization effects are expected to be greatly reduced and likely to be unimportant \citep{Dudik_2017}.

\section{Discussion and Conclusions}
We used observations of a major flare to test whether the plasma structure of the transition region is compatible with classical loop models in which thermal conductivity is assumed to be well described by the Spitzer-Harm collisional theory \citep{Spitzer_Harm_1953}. According to these loop models, the ratio of the emission measures at two different temperatures, as derived from observations of a loop footpoint, should be roughly equal to the ratio of the temperatures \citep{Emslie_Bradshaw_2022}. In our study, the EM ratio between \ion{Si}{4} 1402.770\AA, and AIA 171\AA\ provides the critical test for the predicted thermal structure. We note from Figure~\ref{Figure_em_ratios},h  that in the flare regions, the EM ratio is smaller than the theoretical value ($T_{171}/T_{1402}\,=\, 10$) by two to three orders of magnitudes, which is well outside any possible sources of observational uncertainty. In non-flare regions, (region 1) the ratio is the order of unity, still lower than the expected value, but in agreement with many other observational studies. 

The ratio of  \ion{Fe}{21} 1354.066\AA\ over AIA 171\AA\ can be sampled only at the footpoint of a hot loop (region 3). This EM ratio, seen in Fig.~\ref{Figure_em_ratios}i is in the range of 50 to 200 which is close to the theoretical range ($T_{1354}/T_{1402}\simeq\, 140$). Therefore, the upper transition region of the post-flare hot loops does seem to be compatible with the Spitzer-Harm thermal conduction. Again, this finding is in good agreement with many other studies.

In the previous section we considered several observational effects that may have influenced our results. The EM ratio of 131\AA\ over \ion{Fe}{21} 1354.066\AA\ is roughly equal to $1$, as expected because both emissions sample the 10~MK plasma.  This expected result indicates that the co-alignment, the filling factor and the instrument calibration are not serious issues for our study. Moreover, the abundance selection does not affect the radiations originating from silicon and iron elements as they are both low-FIP species. The influence of abundance is possibly visible in the EM ratio of \ion{C}{2} 1334.535\AA\ over \ion{Si}{4} 1402.770\AA\ as the ratio changes between the different regions as if one passes from coronal to photospheric abundances. Finally, since we measure the emission measure ratios tens of minutes after flare onset, it is unlikely that non-Maxwellian effects are playing a major role in the line formation.

We conclude, therefore, that the observations and their interpretations are in fact correct and that the egregious  disagreement between the observations and the TR models lies with the theory. As discussed in the Introduction, the form of the DEM in the TR for the loop models follows directly from the form of the Spitzer-Harm thermal conductivity, because it determines the temperature scale height as inferred from Equation~\ref{eqn:H_T}. The observations imply that the actual temperature scale height in the lower TR is orders of magnitude larger that derived from Equation~\ref{eqn:H_T}. There are only two possibilities for increasing the scale height in Equation~\ref{eqn:H_T}, either the heat flux is much smaller than expected or the coefficient of thermal conduction is much larger.

The first possibility has been analyzed in depth by Emslie and co-workers \citep{Bian16,Emslie18,Allred22,Emslie_Bradshaw_2022}. These authors have argued that in the hot coronal sections of flare and active region loops, the electron mean free path becomes so large that turbulence is generated. This turbulence acts to scatter the particles, reducing the effective mean free path and the effective heat flux. In the transition region the mean free path is small due to the high densities, so Spitzer-Harm still holds there, but the temperature structure of the whole loop is greatly smoothed by the effect of the heat flux reduction at high temperatures \citep{Emslie_Bradshaw_2022}. Observational support for these ideas can be found in the large non-thermal widths commonly measured for flare emission lines \citep[e.g.][]{Landi03,Young_2015,Reeves20} and for the substantially longer than expected cooling times seen for flare loops \citep[e.g.][]{Ryan13}. The turbulent suppression of flare cooling can naturally account for both these observations \citep{Allred22}.  \citet{Emslie_Bradshaw_2022} have also examined the effect of coronal turbulence on the DEM of hot active region loops, and concluded that the suppression of the heat flux is sufficient to account for the large emission measure at low temperatures. Note that the DEM curve shown in Figure 4 of that paper resembles the curves shown in Figure~\ref{EMasafuncoftemp} above. On the other hand, many observational studies of impulsively heated, but non-flare, active region loops have found that even though they also disagree with standard theory for the DEM of the lower TR, their cooling profiles are well described by Spitzer-Harm thermal conduction \citep[e.g.,][]{Klimchuk08,Viall15}. The model of \citet{Emslie_Bradshaw_2022} seems promising, but clearly more work needs to be done on detailed comparison with observations.

The second possibility for resolving the lower TR problem is that the coefficient of thermal conductivity there must be orders of magnitude larger than the Spitzer-Harm value. To our knowledge, there have been only two studies, and none recent, proposing the enhancement of thermal transport at low temperatures. One is the work of \citet{Shoub_1983,Shoub_1987} who argued that since the electron mean free path varies as $u^4$, where $u$ is the particle velocity, then somewhat nonthermal particles may have a large effect on the heat flux and on the particle distribution functions. Given the high densities  in the lower TR of flare loops, generally $10^{12}$ cm$^{-3}$ or so, this hypothesis needs much further study in order to determine whether these effects can increase the effective mean free path of the heat-carrying particles, primarily electrons, sufficiently to account for our observations. The other hypothesis is that true hydrodynamic-like turbulence kicks in at low temperature, and that this turbulence accounts for the energy transport rather than thermal conduction \citep{Cally90}. There is no doubt that under many circumstances turbulence can transport heat much more effectively than thermal conduction, for example, in the solar convection zone, but it is far from clear whether such turbulence can ever be generated in the low-beta corona and transition region \citep[e.g.][]{Klimchuk21}. It appears, therefore, that the physical origin of the form of the DEM of the lower TR is still not fully understood.

\section*{acknowledgements}
P.R.~Young  acknowledges support from the GSFC Internal Scientist Funding Model
competitive work package program and the Heliophysics Guest Investigator program. S.K. ~Antiochos acknowledges support from the NASA LWS Program, Grant 80NSSC22K0892 to the University of Michigan.
CHIANTI is a collaborative project involving George Mason University, the University of Michigan (USA), University of Cambridge (UK) and NASA Goddard Space Flight Center (USA).

\bibliography{bibliography_transition_region_leaf_2023}{}
\bibliographystyle{aasjournal}

\end{document}